\documentclass[twocolumn]{article}
\usepackage[fontsize=9pt]{scrextend}
\usepackage[T1]{fontenc}
\usepackage[english]{babel}
\usepackage{amssymb}
\usepackage{lipsum, color}
\usepackage{graphicx}
\usepackage{ulem}
\usepackage{xcolor}
\usepackage{pdfpages}
\usepackage{mathtools}
\usepackage[utf8]{inputenc}
\usepackage{authblk}
\usepackage{upgreek}
\usepackage{pdfpages}
\usepackage[a4paper, total={6.5in, 8.5in}]{geometry}

\usepackage[T1]{fontenc}       

\makeatletter
\let\@fnsymbol\@arabic
\makeatother

\title{\textbf{Scale-invariant biomarker discovery in urine and plasma metabolite fingerprints}}
\author[a,*]{Helena U. Zacharias\thanks{helena.zacharias@ukr.de}}
\author[b,*]{Thorsten Rehberg}
\author[b]{Sebastian Mehrl}
\author[c]{Daniel Richtmann}
\author[c]{Tilo Wettig}
\author[a]{Peter J. Oefner}
\author[b]{Rainer Spang}
\author[a]{Wolfram Gronwald\thanks{wolfram.gronwald@ukr.de}}
\author[b]{Michael Altenbuchinger\thanks{michael.altenbuchinger@ukr.de}}
\affil[a]{\textit{Institute of Functional Genomics, University of Regensburg, Regensburg, Germany}}
\affil[b]{\textit{Statistical Bioinformatics, Institute of Functional Genomics, University of Regensburg, Regensburg, Germany}}
\affil[c]{\textit{Department of Physics, University of Regensburg, Regensburg, Germany}}
\affil[*]{\footnotesize{The authors wish it to be known that, in their opinion, the first two authors should be regarded as Joint First Authors.}}

\date{}

\begin{document}
\maketitle

\begin{abstract}

\quad\textbf{Motivation:}
Metabolomics data is typically scaled to a common reference like a constant
volume of body fluid, a constant creatinine level, or a constant area under the 
spectrum. Such normalization of the data, however, may affect the selection of biomarkers and the biological interpretation of results in unforeseen ways.

\textbf{Results:} 
First, we study how the outcome of hypothesis tests for differential 
metabolite concentration is affected by the choice of scale. Furthermore, we observe this interdependence also for different classification approaches.
Second, to overcome this problem and establish a scale-invariant biomarker discovery algorithm, we extend linear zero-sum 
regression to the logistic regression framework and show in two applications
to ${}^1$H NMR-based metabolomics data how this approach overcomes the scaling problem. 

\textbf{Availability:} 
Logistic zero-sum regression is available as an \textit{R} package as well as a high-performance computing implementation that can be  downloaded at https://github.com/rehbergT/zeroSum.

\end{abstract}

\section{Introduction}

Metabolomics is the comprehensive study of all small organic compounds in a biological specimen \cite{German2005}.
Metabolite concentrations in body fluids such as  urine, serum, and plasma have proven valuable in predicting disease onset and 
progression \cite{Dawiskibaetal2014, Gronwaldetal2011, Zachariasetal2013a, Zachariasetal2015, Elliottetal2015}. 
Metabolomic data can be generated by a variety of methods of which mass spectrometry and ${}^1$H NMR spectroscopy 
are the most common.
While its high sensitivity makes mass spectrometry the preferred method for discovery projects,
the high reproducibility of NMR data \cite{Keun_2002} is ideal for applications in precision medicine. 

${}^1$H NMR allows for the simultaneous detection of all proton-containing metabolites present at sufficient concentrations in biological specimens. 
Furthermore, NMR signal volume scales linearly with concentration. The complete set of NMR signals acquired from a given specimen is called its 
metabolite ``fingerprint''. Due to differences in pH, salt concentration, and/or temperature, relative displacement of signal positions occurs 
across spectra. Binning schemes can compensate for this effect by splitting spectra into segments called bins and summing up signal volumes contained therein. Equal-sized bins are commonly used, albeit other schemes such as adaptive binning 
have been suggested \cite{DeMeyeretal2008, Andersonetal2011}. Binned fingerprint data is the typical starting point for subsequent multivariate data analysis
including biomarker discovery and specimen classification using methods such as 
Support Vector Machines \cite{Cortesetal1995,Burgesetal1998},  ridge \cite{Hoerl1970}, and LASSO regression \cite{Tibshirani1996}.

Metabolite fingerprints need to be scaled to a common unit. Typical examples include mmol metabolite per ml plasma, 
mmol metabolite per mmol creatinine in urine, or the relative contribution of a bucket to the total
spectral intensity of an NMR spectrum. In practice this is achieved by dividing each spectrum by the unit 
defining quantity: the intensity of an NMR reference such as TSP, the intensity of creatinine, or
the total spectral intensity. This scaling of the raw data defines the measurement, but it also serves a 
second purpose: scaling corrects for unwanted experimental and physiological variability in the raw spectra. 
Correcting unwanted variability is usually called normalization. 
Instrument performance can vary and so can patient-specific 
parameters like the fluid balance, which is affected by drinking, respiration, defecation, perspiration, or 
medication, thus altering urine metabolite concentrations without reflecting disease state 
\cite{Ryanetal2011, Craigetal2006}. 

All scales have their pros and cons: the NMR reference can normalize for spectrometer performance \cite{Viantetal2005} but not 
for unwanted variability in urine density. Choosing creatinine as a standard for urine assumes the absence of inter-individual differences in the production and renal excretion of creatinine \cite{Waikaretal2010}. In fact, creatinine production and
excretion is affected by sex, age, muscle mass, diet, pregnancy, and, most importantly, renal pathology 
\cite{Stevensetal2009, Curhan2005}. Normalization to a constant total spectral intensity or spectral mapping to a reference 
spectrum assume that the total amount of metabolites is constant over time and across patients and that 
spectra are not contaminated by signals that do not represent metabolites. However, this is not always the case 
\cite{Hochreinetal2015, Rossetal2007}.
In fact, for urinary specimens of patients suffering from proteinuria the additional protein signals greatly increase total 
spectral area, and scaling to a constant total intensity
systematically underestimates metabolite abundances. Similarly, excessive 
glucose uptake, for example by a glucose infusion, leads to high total spectral intensities that are dominated by glucose and its metabolites.
While the high values for these metabolites correctly reflect the metabolic state for these patients,
their influence on the total spectral intensity leads to systematic underestimation of metabolites not related 
to glucose metabolism.

In general, the preferred scaling protocol depends on the specific data set to be investigated. However, for some data sets 
it is not possible to use the same scale for all patients in the cohort. We will describe such data sets below. In this case different protocols need to be used for different 
patients, introducing new challenges to data analysis.

In this contribution, we first studied how the choice of scale affects statistical analysis, the selection of 
biomarkers, and patients' diagnosis by these biomarkers. 
We report on  two supervised metabolomics data analysis scenarios, namely urine and plasma biomarker discovery in 1D ${}^1$H NMR metabolite fingerprints for the early detection of acute kidney injury onset after cardiac surgery 
\cite{Zachariasetal2013a, Zachariasetal2015}. 
In both applications we tested metabolites for differential abundance using alternative scaling protocols. 
We observed pronounced disagreements between the lists of significantly differential metabolites depending 
on how the same data was scaled. More importantly, the different scalings led to inconsistencies in the classification of individual patients. 
In view of these observations, reproducibility of metabolic studies is only possible if the exact same scaling 
protocols are used.

To overcome this problem, we extended zero-sum regression \cite{Linetal2014,Altenbuchingeretal2016}, which has recently been demonstrated to be invariant under any rescaling of data \cite{Altenbuchingeretal2016}, to logistic zero-sum regression and compared it to 
two standard methods for constructing multivariate signatures: LASSO logistic regression and Support Vector 
Machines in combination with \textit{t}-score based feature filtering. Unlike the latter methods, logistic zero-sum 
regression always identifies the same biomarkers regardless of the scaling method. Consequently, prior data normalization may be omitted completely.    
We make logistic zero-sum regression available as an \textit{R} package and as a high-performance computing software that can be downloaded at https://github.com/rehbergT/zeroSum.

\section{Experimental Section}

\subsection{Data Sets}
The first data set comprised 1D ${}^1$H NMR fingerprints of $n$ = 106 urine specimens, which had been collected from patients 24 h after cardiac surgery 
with cardiopulmonary bypass (CPB) use at the University Clinic of Erlangen \cite{Zachariasetal2013a}. Of these 106, 34 were diagnosed with 
acute kidney injury (AKI) 48 h after surgery. The challenge in this data is to define a urinary biomarker  signature that allows for
the early detection of AKI onset (Acute Kidney Injury Network (AKIN) stages 1 to 3).
The second data set consisted of 85 EDTA-plasma specimens, which had been collected 24 h 
post-op from a subcohort of the original cohort of 106 patients undergoing cardiac surgery with CPB use and which had been subjected to 10 kDa cutoff filtration \cite{Zachariasetal2015}. 
In total, 33 patients out of these 85 patients were diagnosed with postoperative AKI \cite{Zachariasetal2015}. Again our goal is to detect biomarkers for an earlier detection of AKI.

\subsection{NMR Spectroscopy}
A total of 400 $\upmu$l of urine or EDTA-plasma ultrafiltrate was mixed with 200 $\upmu$l of phosphate buffer, pH 7.4, and 50 $\upmu$l of 0.75\% (w) 3-trimethylsilyl-2,2,3,3-tetradeuteropropionate (TSP) dissolved in deuterium oxide as
the internal standard (Sigma-Aldrich, Taufkirchen, Germany). NMR experiments were carried out on a 600 MHz Bruker Avance III (Bruker BioSpin GmbH, Rheinstetten, Germany) employing a triple resonance (${}^1$H, ${}^{13}$C, ${}^{31}$P, ${}^2$H lock) cryogenic probe equipped with $z$-gradients and an automatic cooled sample changer. For each sample, a 1D ${}^1$H NMR spectrum was acquired employing a 1D nuclear Overhauser enhancement spectroscopy (NOESY) pulse sequence with solvent signal suppression by presaturation during relaxation and mixing time following established protocols \cite{Zachariasetal2013b, Gronwaldetal2008}. NMR signals were identified by comparison with reference spectra of pure compounds acquired under equal experimental conditions \cite{Zachariasetal2013b}.

\subsection{Data extraction}
 
The spectral region from 9.5 to $-$0.5 ppm of the 1D spectra was exported as even bins of 0.001 ppm width employing Amix 3.9.13 (Bruker BioSpin). 
The data matrix was imported into the statistical analysis software \textit{R} version 3.3.2. For the urinary spectra, the region 
 6.5 -- 4.5 ppm, which contains the broad urea and water signals, was excluded prior to further analysis. For plasma spectra, the region 6.2 -- 4.6 ppm, containing the urea and remaining water signals, was removed prior to analysis. In addition, the regions 3.82 -- 3.76 ppm, 3.68 -- 3.52 ppm, 3.23 -- 3.2 ppm, and 0.75 -- 0.72 ppm, corresponding to filter residues and free EDTA, were excluded prior to classification for the plasma specimens.

\subsection{Scaling and normalization}
We compare our scaling- and normalization-invariant approach to standard analysis strategies that were applied to data preprocessed by four state-of-the-art normalization protocols. For creatinine normalization of urinary data we   
divided each bucket intensity by the summed intensities of the creatinine reference region ranging from 3.055 to 3.013 ppm of the corresponding 
NMR spectrum.
When scaling to the total spectral intensity we summed over all signal intensities from 9.5 to 0.5 ppm after exclusion of the water and 
urea signals and divided each bucket intensity by this sum.
When scaling to the signal intensity of the NMR reference compound, in our case TSP, we summed up the intensities of the TSP buckets ranging 
from $-$0.025 to 0.025 ppm and divided each bucket intensity by this sum. This normalization method corrects for differences in 
spectrometer performance \cite{Zachariasetal2015}, but not for differences in fluid intake.
Finally, we also evaluated Probabilistic Quotient Normalization (PQN) \cite{Dieterleetal2006}.
PQN follows the rationale that changes in concentration of one or a few metabolites affect only small segments of the spectra, whereas specimen dilution, for example due to differences in fluid intake in case of urinary specimens, influences all spectral signals simultaneously \cite{Dieterleetal2006}. 
Following \cite{Dieterleetal2006} we first normalized each spectrum to its total spectral intensity. Then, we took the median across all these spectra as the reference spectrum and calculated 
the ratio of all bucket intensities between raw and reference spectra. The median of these ratios in a sample was our final 
division factor. 

For subsequent analysis, only the spectral region between 9.5 and 0.5 ppm was taken into account. To compensate for slight shifts in signal positions across spectra due to small variations in sample pH, salt concentration, and/or temperature,  bucket intensities across ten buckets were fused together in one bucket of 0.01 ppm width 
by summing up the individual bucket intensities.
The resulting intensities were $\log{}_2$ transformed to correct for heteroscedasticity, and all subsequent data analysis was performed 
with these values. 
\begin{figure*}[!t]
\includegraphics[width=1\textwidth]{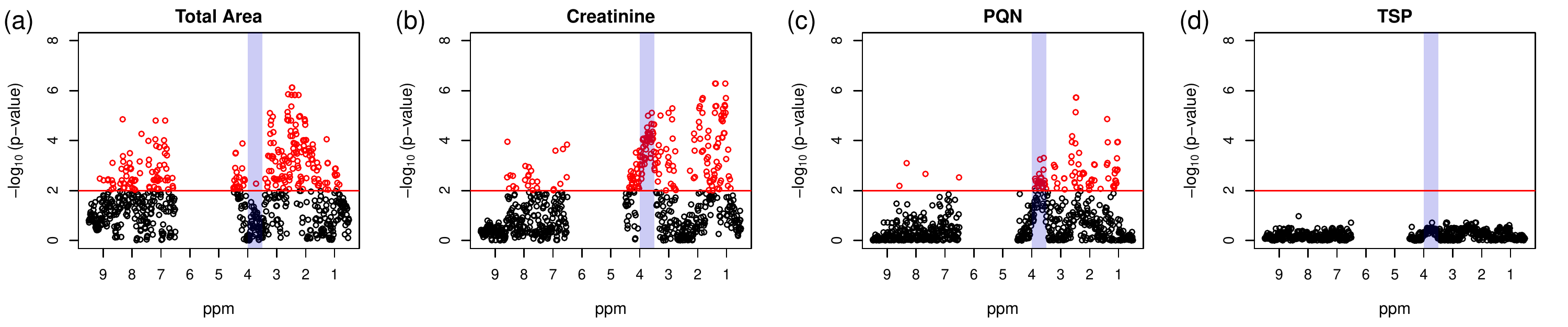}
\includegraphics[width=1\textwidth]{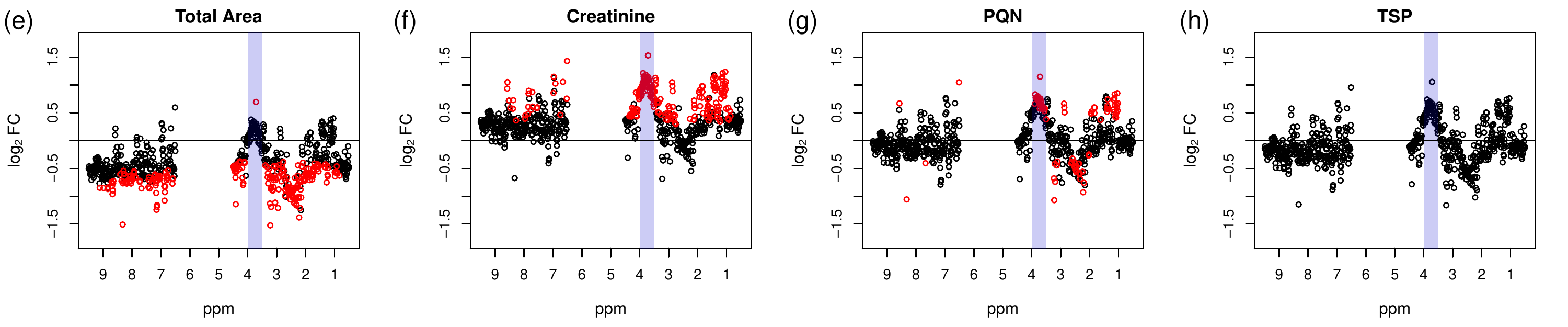}
\caption{\label{t-test}$-\log_{10}(\textrm{$p$-values})$ of moderated $t$-test analysis comparing healthy versus diseased patients for the urinary AKI data set after preprocessing with four different normalization methods, i.e., scaling to (a) equal total spectral area, (b) scaling to creatinine, (c) PQN, and (d) scaling to TSP, respectively, plotted versus the ppm regions of the corresponding NMR features (upper figure). A red line marks the significance level for Benjamini-Hochberg (B/H) adjusted $p$-values below 0.01, corresponding to a false discovery rate (FDR) below 1$\%$. All NMR features with a B/H-adjusted $p$-value below 0.01 are represented as red dots. The lower figures, (e - h), show the corresponding $\log{}_2$ fold changes ($\log{}_2$ FC) versus the ppm regions of the corresponding NMR features. $\log{}_2$ FCs were calculated as AKI minus non-AKI, thus positive $\log{}_2$ FCs correspond to higher values in AKI than in non-AKI.}
\end{figure*}

\subsection{Classification algorithms}
We compared logistic zero-sum regression to two classification algorithms for normalized data, namely a Support Vector Machine \cite{Cortesetal1995} employing a linear kernel function
combined with univariate feature filtering (f-SVM) and the least absolute shrinkage and selection operator (LASSO) logistic regression 
\cite{Tibshirani1996}. In the LASSO logistic regression and the zero-sum models, the penalizing parameter $\lambda$ that balances the bias variance trade-off was
optimized in an internal cross-validation. LASSO models were trained utilizing the \textit{R} package \textit{glmnet} \cite{Friedmanetal2010}. The zero-sum models were trained by employing our \textit{zeroSum} software, which is implemented in two versions: first, a version optimized for high-performance computing (HPC), and second, an \textit{R} package for ordinary desktop computers. We utilized the \textit{zeroSum} HPC implementation throughout the article. 

The parameter $\lambda$ induces sparseness of the models and circumvents over-fitting for LASSO and zero-sum models. Similarly, we optimized the parameters of f-SVM. This was done in a two-fold nested cross-validation, i.e.,\ in each cross-validation step we screened for the best feature threshold, allowing for $\{2^1,2^2,2^3,2^4, 2^5, 2^6\}$ included features, and the matching best cost parameter, which was screened in $\{2^{-5},2^{-4},\ldots, 2^4, 2^5\}$. For training the SVM we utilized the \textit{R} package \textit{e1071} \cite{Dimitriadouetal2011}.

\section{Results} 
\subsection{Stand-alone biomarker discovery depends on the choice of scale}
\label{SigTest}

Metabolic biomarkers can be metabolites or just spectral features.  
Moreover, they can be identified as stand-alone predictors or as part of a multivariate biomarker signature. 
Before we focus on signatures we study the effect of scales on stand-alone biomarkers.
More precisely, we focus on spectral buckets with differential intensities between two classes of samples, i.e., from patients who developed acute kidney injury after cardiac surgery versus from those that did not.  

Bin intensities were normalized to (a) a constant total spectral intensity, 
(b) a constant creatinine intensity, (c) a median reference spectrum employing the most probable quotient and (d) a constant 
intensity of the NMR reference. For scoring normalized bin intensities as potential biomarkers we used Benjamini-Hochberg (B/H) corrected $p$-values 
from the moderated \textit{t}-statistics implemented in the \textit{R} package LIMMA \cite{Ritchieetal2015} and $\log{}_2$ fold changes between 
the classes.

Figure \ref{t-test} shows plots of spectral bin positions against $-\log{}_{10}$($p$-values) (a-d) and $\log{}_2$ fold changes (e-h) between AKI and non-AKI urine specimens. From left to right the plots correspond to (a,e) normalization by total spectral intensity, (b,f) normalization to creatinine, (c,g) PQN, 
and (d,h) normalization to TSP.   
The red line in plots (a-d) marks a significance level of 0.01, corresponding to a false discovery rate (FDR) below 1$\%$. 
Significant buckets are plotted in red in all eight plots.

{\bf Normalization to total spectral intensity:} On this scale we observed almost exclusively negative fold changes indicating 
lower bucket intensities for AKI in comparison to non-AKI. 242 features were significant with 
features corresponding to carnitine, 2-oxoglutaric acid, and glutamine ranking highest.
Supplementary Table S1 (a) lists raw and B/H-adjusted $p$-values as well as metabolite assignments of the top ten significant buckets.

\begin{figure}[ht]
\includegraphics[width=0.5\textwidth]{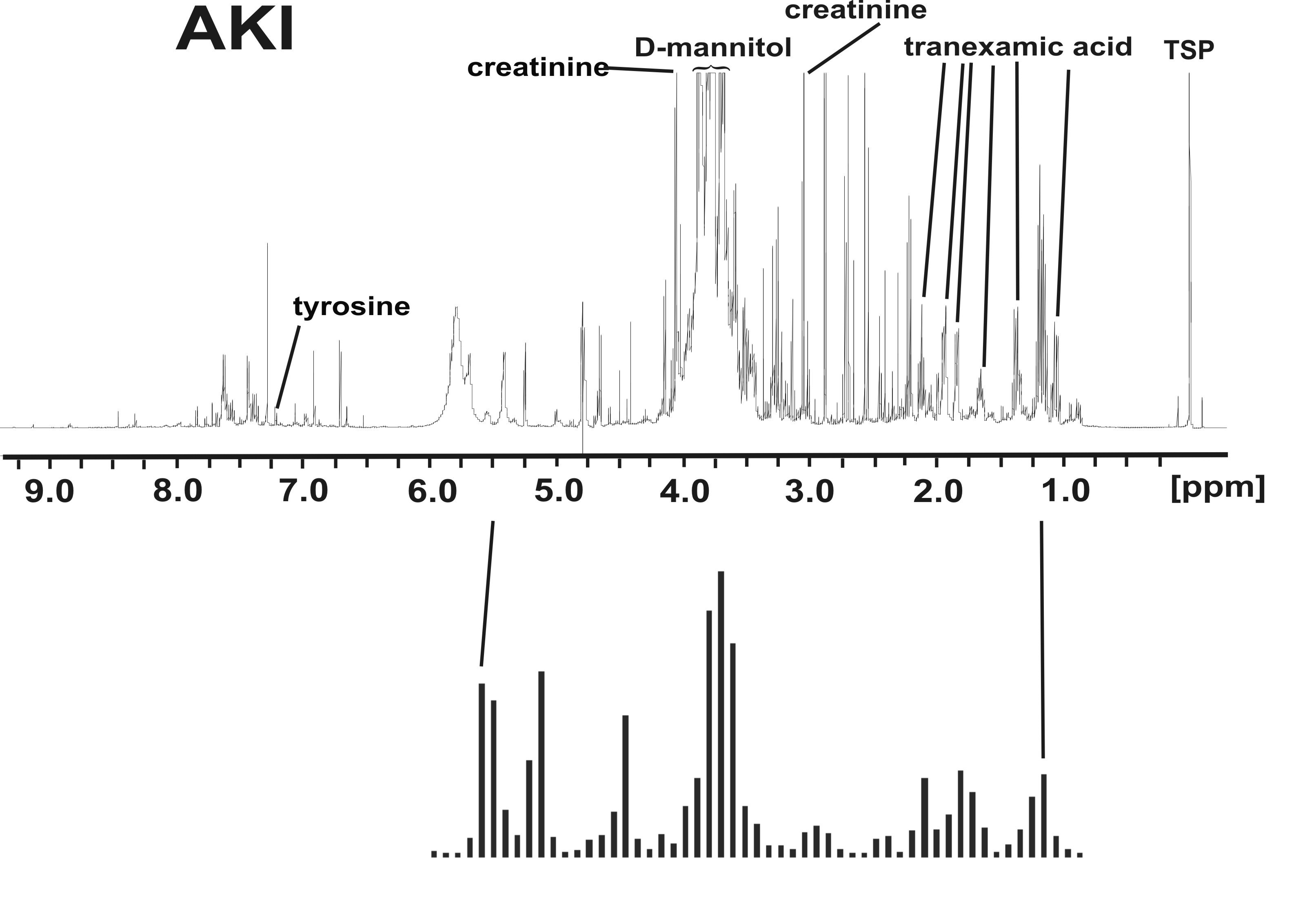}
\caption{\label{AKI_urin}Exemplary 1D ${}^1$H NMR spectrum of a urine specimen collected 24 h after surgery from an AKI patient. The lower histogram illustrates the common binning procedure.}
\end{figure}

Technical artifacts of the scale become apparent too:
the spectral region from 4 to 3.5 ppm, highlighted as a blue band, hardly comprised any significant features. 
This region is dominated by signals from sugars such as D-mannitol, which had been used as a pre-filling material 
for the tubes of the CPB machine \cite{Zachariasetal2013a}. As D-mannitol exhibits a rather 
large number of NMR signals, as highlighted in an exemplary urine spectrum shown in Figure \ref{AKI_urin}, it 
comprised between $15\%$ and $82\%$ of the total spectral areas.
On a scale that makes the total spectral area the same across all samples these signals cannot serve as biomarkers.
However, all patients take up D-mannitol during surgery, and the amount of D-mannitol still found in urine 24 h past surgery is 
modulated by actual kidney function. D-mannitol entangles kidney function with the total spectral area.
Therefore, the spectral area can not be recommended for scaling biomarkers, at least not in this context.

{\bf Normalization to creatinine:} For this scale we observed different analysis results.
$\log{}_2$ fold changes were predominantly positive indicating higher metabolite levels in AKI patients.
204 buckets were significant, and the top ten are listed in Supplementary Table S1 (b). The biomarker ranked highest was tranexamic acid, which is given in cardiac surgery to prevent excessive blood loss. 
Strikingly, the region between 4 and 3.5 ppm (blue band) containing the D-mannitol signals was highly significant on this scale. 

However, this scale confounds with aspects of kidney function.
In response to kidney disease urinary creatinine excretion rates can be affected \cite{Waikaretal2010}. Thus, normalization 
to creatinine can also obscure other metabolite biomarkers if their excretion correlates with that of creatinine. 

{\bf Probabilistic Quotient Normalization:} On this scale we 
observed both positive and negative fold changes in almost equal numbers. Only 89 features were now significant, and the top ten are listed in Table S1 (c), with carnitine and glutamine 
as the leading biomarkers. Again, the blue shaded region now covers significant features, although the number is much lower than after creatinine scaling. 

Due to the strong conceptual similarity between PQN and normalization to total spectral intensity, the D-mannitol artifact
also compromises the use of PQN. 

{\bf TSP normalization:} On this scale we did not detect any significant biomarkers.
The scaling method corrects only for differences in spectrometer performance, not for changes in global metabolite concentration.
Due to large variability in urine density throughout the cohort, it can not be used in this context.

\begin{figure}[ht]
\centering
\includegraphics[width=0.24\textwidth]{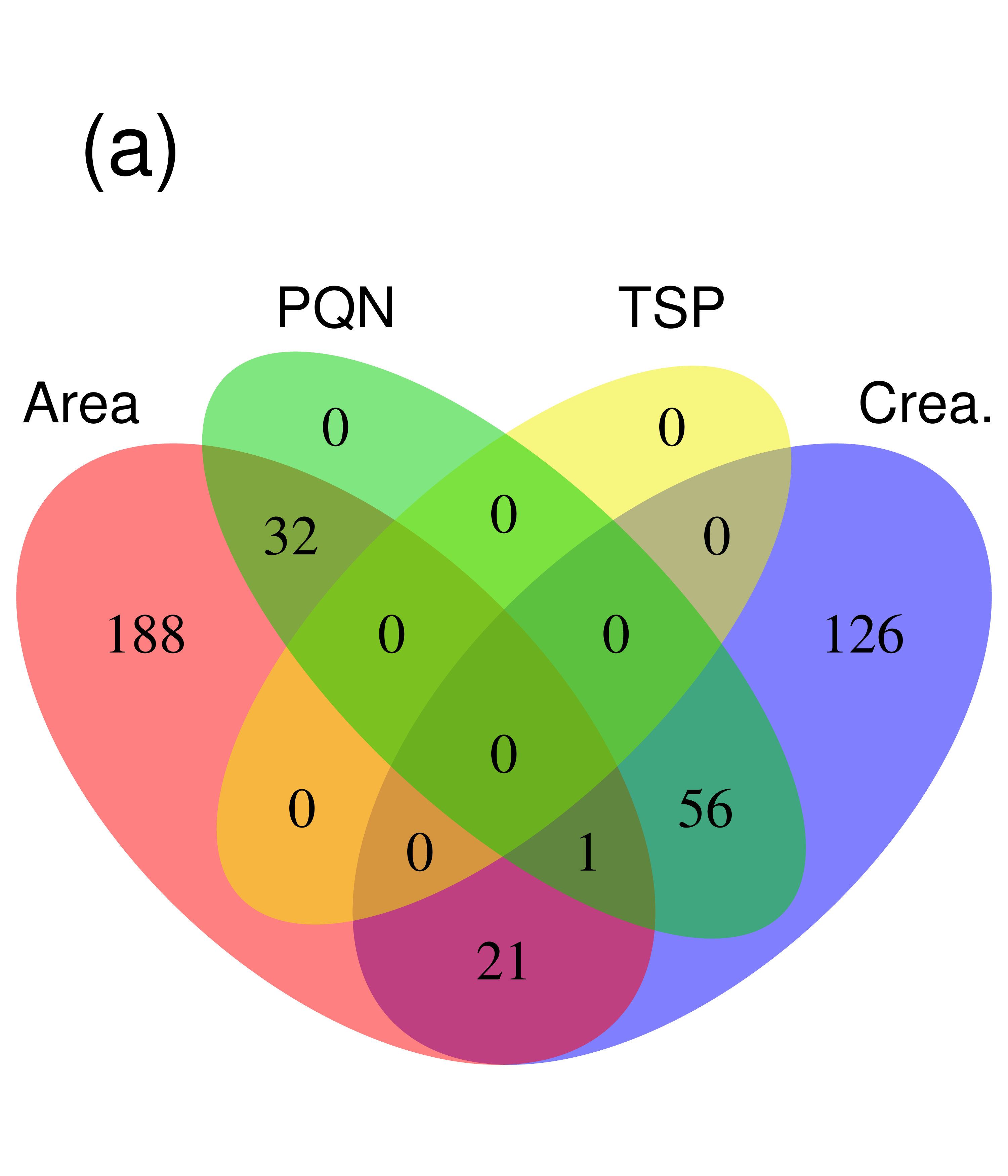}
\includegraphics[width=0.24\textwidth]{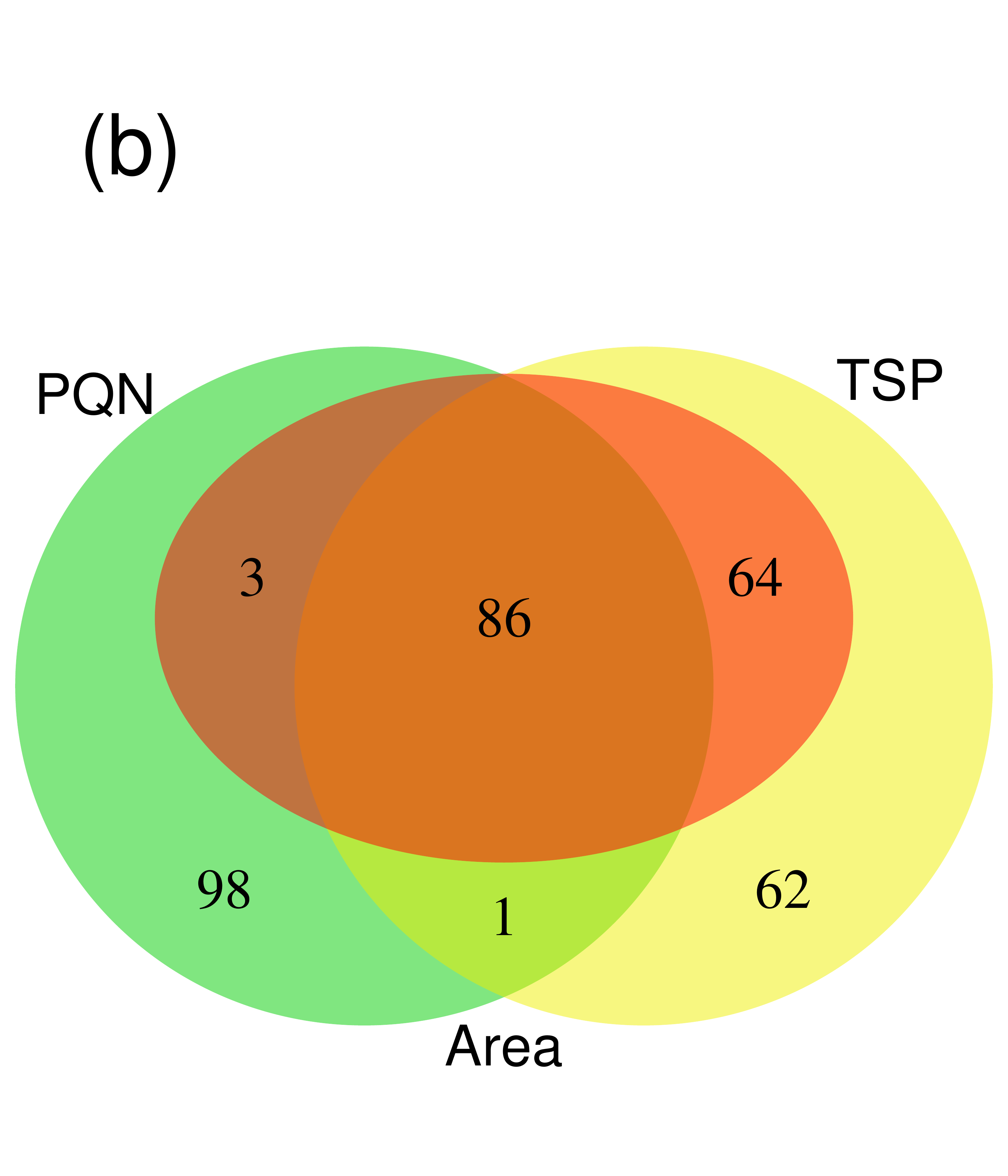}
\caption{\label{Venn_AKI_classif} Number of significant buckets and their overlap between the different scaling methods for (a) the AKI urine and (b) the AKI plasma data set.}
\end{figure}

We observed only one bucket at 3.715 ppm, identified as an overlap of propofol-glucuronide, broad protein signals, and 
tentatively D-glucuronic acid, which obtained a significant $p$-value on three scales (Figure \ref{Venn_AKI_classif}a).

The plasma data biomarker discovery was not consistent across scales either
(Supplementary Table S2, Figure S1, and Figure \ref{Venn_AKI_classif}b). Scaling to total spectral area and scaling to TSP predominantly identified metabolites that accumulate in the 
blood of patients developing AKI. One might explain this observation by a reduced glomerular filtration in these 
patients. However, the PQN data immediately challenged this interpretation, as it identified a large 
set of down-regulated metabolites in AKI. Creatinine normalization is not common for plasma metabolomics and was 
not investigated here.

\subsection{Multivariate biomarker signatures can be scale dependent} \label{Classif_bias}
Biomarkers can be combined to biomarker signatures in multivariate analysis. Machine-learning algorithms are used to learn these signatures from training data. 
We tested two of these algorithms, namely a linear SVM with $t$-score based feature filtering (f-SVM) and standard logistic LASSO regression.
Both methods implement linear signatures of the form
\begin{equation}
   \beta_0+\sum_{j=1}^p\beta_j\,\log_2(\Gamma_i\,X_{ij}),
\label{regshift}
\end{equation}
where $X_{ij}$ is the raw signal from bucket $j$ in patient $i$, $\beta_j$ the weight of this feature, and $\Gamma_i$ the scaling factor
used to normalize sample $i$. 
Both methods were used to learn signatures from data normalized in four different ways, namely scaling by total spectral area, scaling to creatinine, PQN, and scaling to TSP.
The performance of the algorithms was tested in cross-validation.

\begin{figure}[ht]
\vspace{5mm}
\includegraphics[width=.48\textwidth]{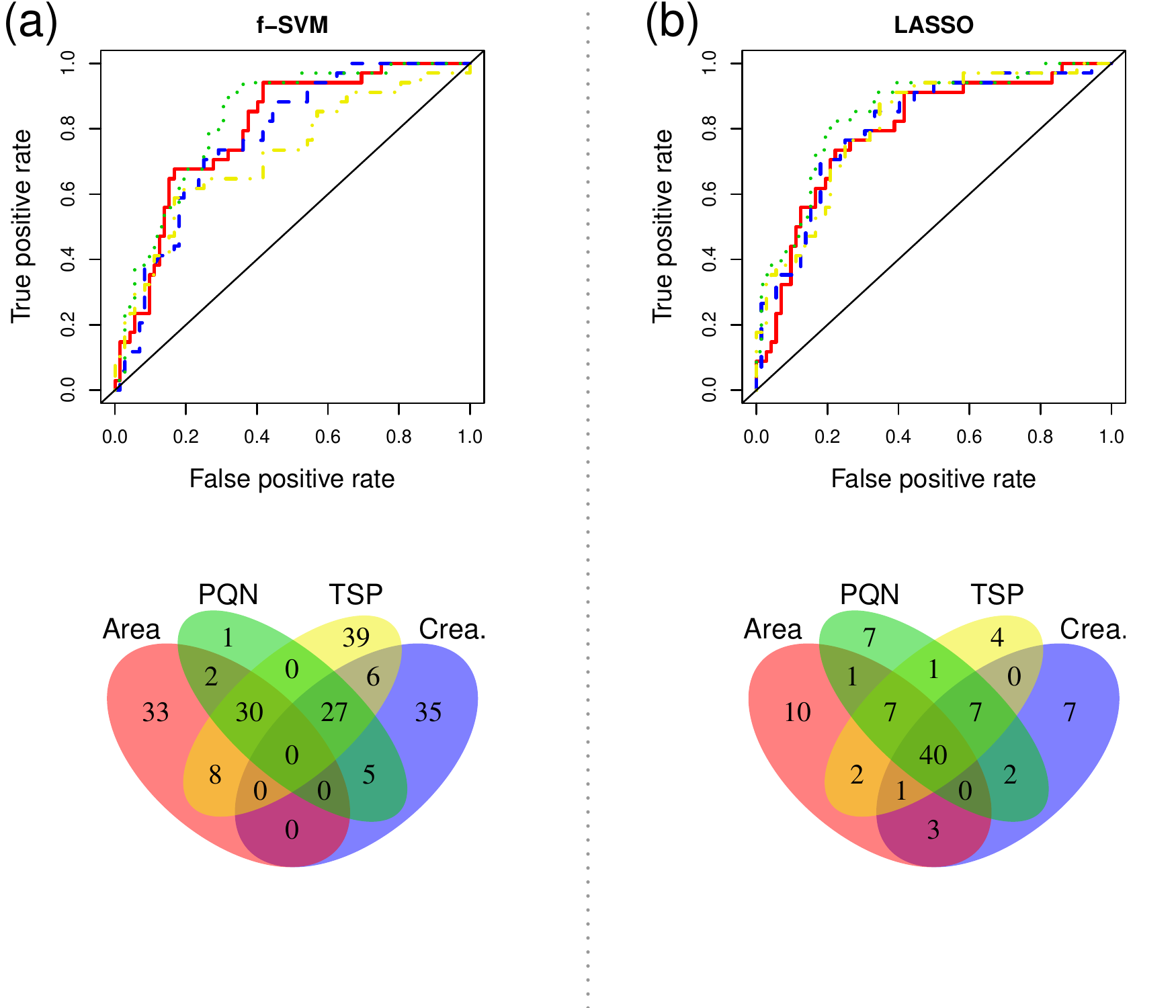}
\caption{\label{ROC_AKI}The urinary AKI data set: Receiver operating characteristic (ROC) curves for two classification approaches, (a) SVM in combination with $t$-test based feature filtering and (b) LASSO, after application of four different normalization strategies: scaling to total spectral area (red solid line), scaling to creatinine (blue dashed line), Probabilistic Quotient Normalization (PQN) (green dotted line), and scaling to TSP (yellow dashed-dotted line). The bottom row shows the number of features included in the respective classification models in Venn diagrams. The corresponding models were built by averaging over all models of the outer CV loop. 
}
\end{figure}

Figure \ref{ROC_AKI}a shows ROC curves of signatures that aim to predict AKI from urinary fingerprints using f-SVM signatures. 
The areas under the ROC curves (AUC-ROC) are summarized in Table \ref{ROC_table}. The higher this area the better the prediction of patient outcome.  
Performances ranged from an AUC of 0.72 for TSP-scaled data to an AUC of 0.83 for PQN data. 
In line with the results for stand-alone biomarkers, the f-SVM algorithm picked different features across scaling methods. Not a single bucket was 
chosen simultaneously for all four scales (Figure \ref{ROC_AKI}a bottom row).
Figure \ref{ROC_AKI}b and Table \ref{ROC_table} show the corresponding results for standard LASSO logistic regression.
The LASSO signatures performed slightly better and were more consistent. 
Nevertheless, there is still a dependence of the chosen biomarkers on the scale. 
Similar results were observed for the plasma data set (Supplementary Figure S2).

\begin{table}[h!]
\caption{\label{ROC_table}AUC-ROC values of three different classification approaches, f-SVM, LASSO logistic regression, and zero-sum regression after application of four different normalization methods, i.e., scaling to total spectral area, scaling to creatinine, Probabilistic Quotient Normalization (PQN), and scaling to TSP, for the (a) urinary AKI and (b) plasma AKI data set.}
\centering
\scriptsize{
\begin{tabular}{| c || c | c | c |}\hline\hline
 \multicolumn{4}{| c |}{\textbf{(a) Urinary AKI data set}}\\
\hline
\textbf{Normalization} & \textbf{f-SVM} & \textbf{LASSO} & \textbf{Zero-sum}\\
\textbf{method} & & \textbf{logistic regr.} & \textbf{logistic regr.}\\
\hline
\textbf{Tot. spec. area} & 0.80 & 0.81 & 0.83\\
\hline 
\textbf{Creatinine} & 0.77 & 0.81 & 0.83\\
\hline
\textbf{PQN} & 0.83 & 0.85 & 0.83\\
\hline 
\textbf{TSP} & 0.72 & 0.79 & 0.83\\
\hline\hline
 \multicolumn{4}{| c |}{\textbf{(b) Plasma AKI data set}}\\
\hline
\textbf{Normalization} & \textbf{f-SVM} & \textbf{LASSO } & \textbf{Zero-sum }\\
\textbf{method} & & \textbf{logistic regr.} & \textbf{logistic regr.}\\
\hline
\textbf{Tot. spec. area} & 0.85 & 0.84 & 0.89\\
\hline
\textbf{PQN} & 0.84 & 0.86 & 0.89\\
\hline
\textbf{TSP} & 0.87 & 0.84 & 0.89\\ 
\hline\hline
\end{tabular}
}
\end{table}
\normalsize

\begin{figure*}
\includegraphics[width=1\textwidth]{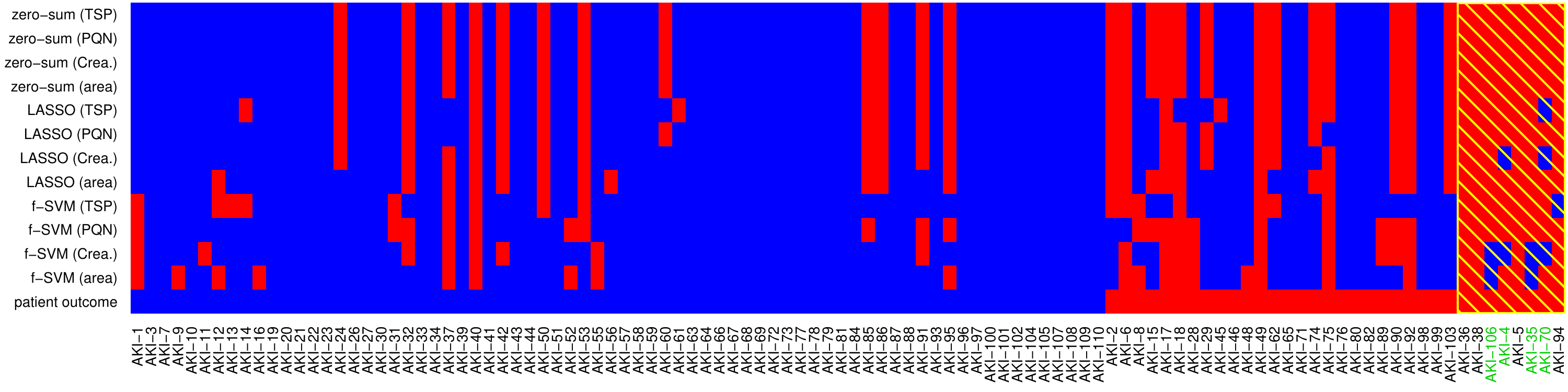}
\caption{\label{classification_summary_AKI}Classification results patient by patient for the urinary AKI data set: The row ``patient outcome'' shows patients that did not develop AKI in blue, patients that developed AKI in red, whereas patients that developed severe AKI (AKIN stage 2 and 3) are further highlighted by the yellow dashed region. Above we give predictions for the onset of AKI for f-SVM, LASSO, and zero-sum, using normalization strategies as indicated in brackets. AKI predictions are shown in red (AKIN stages 1 to 3), while patients predicted as non-AKI are shown in blue. Patients discussed in subsection \ref{scaleDep} are indicated by green sample names.}
\end{figure*}

\subsection{For standard methods prediction of patient outcome is scale dependent}
\label{scaleDep} 

From a clinical perspective, varying biomarkers constitute only a minor problem, as long as they agree in their predictions of outcome. However, they usually do not agree. For the urinary data set, LASSO signatures yielded conflicting predictions in 16\% of patients. For f-SVM signatures, the
percentage increased to 30\%.

Figure  \ref{classification_summary_AKI} summarizes the predictions of AKI onset after cardiac surgery patient by patient for the urinary data set.
The row ``patient outcome'' shows patients in blue who did not develop AKI (AKIN stage 0) and in red those that
developed AKI. Furthermore, the yellow dashed lines highlight
patients with AKIN stage 2 and 3, a more severe manifestation
of kidney injury. The latter patients constitute the high-risk group 
where early detection of AKI onset can save lives \cite{Cruzetal2009}. 
The true outcomes 48 h after surgery are contrasted to the predictions 24 h earlier.
Shown are predictions of f-SVM and LASSO signatures for the four scaling methods. 

We observed that predictions frequently changed with scale and that signatures learned on data that was scaled to, e.g.,  
creatinine did not properly identify the highest risk group. For f-SVM only the signature on PQN data identified all high-risk patients correctly. The LASSO identified high-risk patients for total spectral area and PQN data correctly.

\begin{figure}[ht]
\includegraphics[width=.242\textwidth]{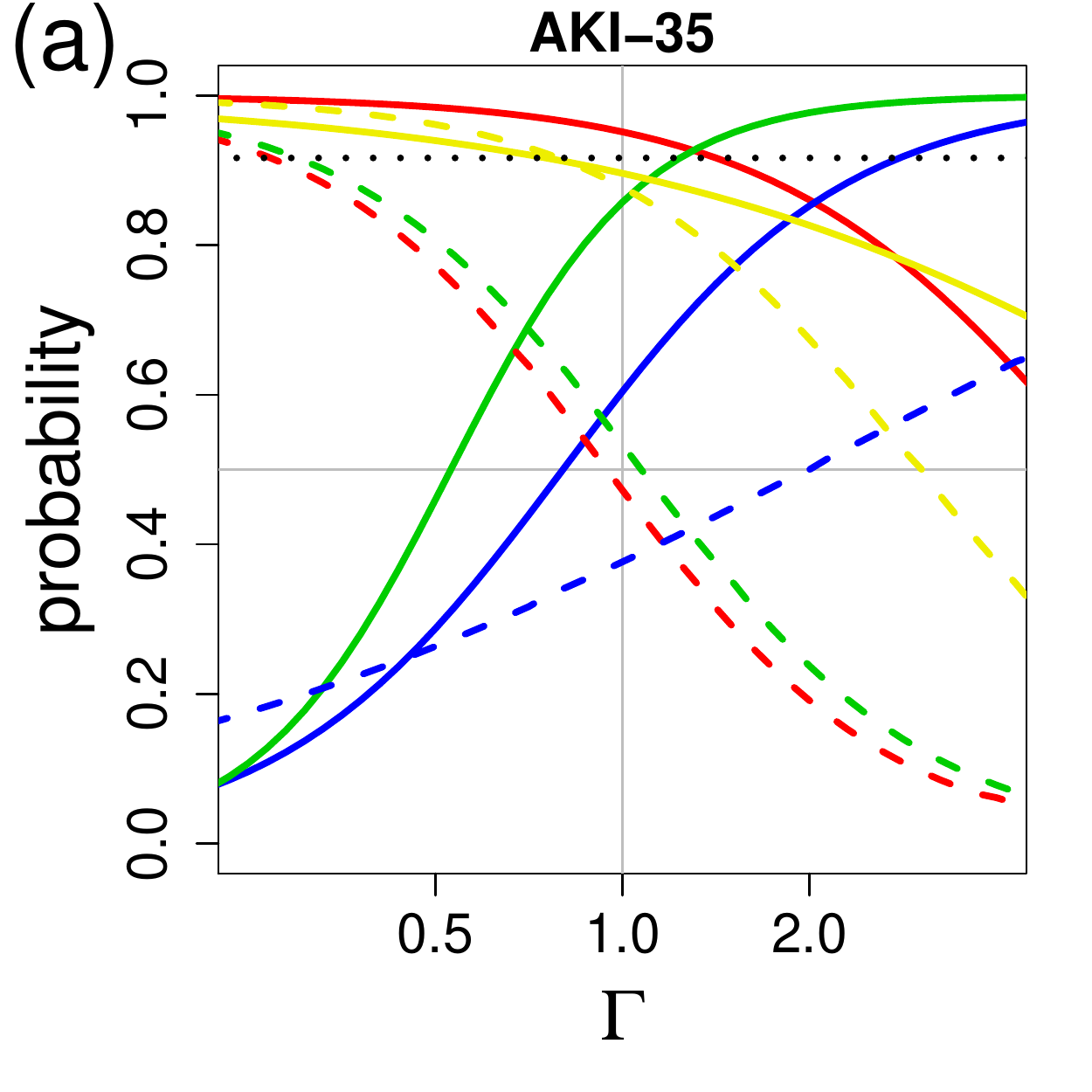}
\includegraphics[width=.242\textwidth]{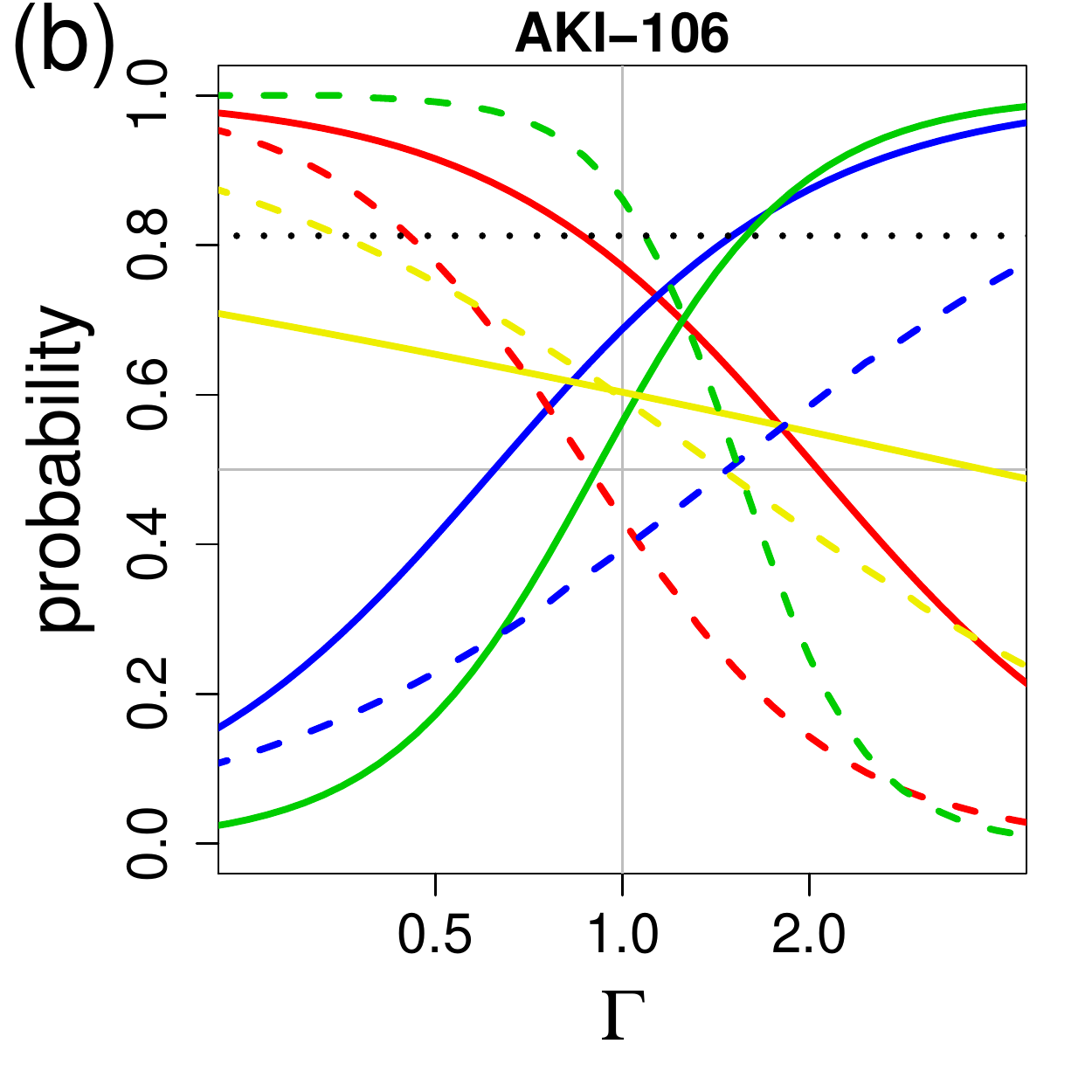}
\includegraphics[width=.242\textwidth]{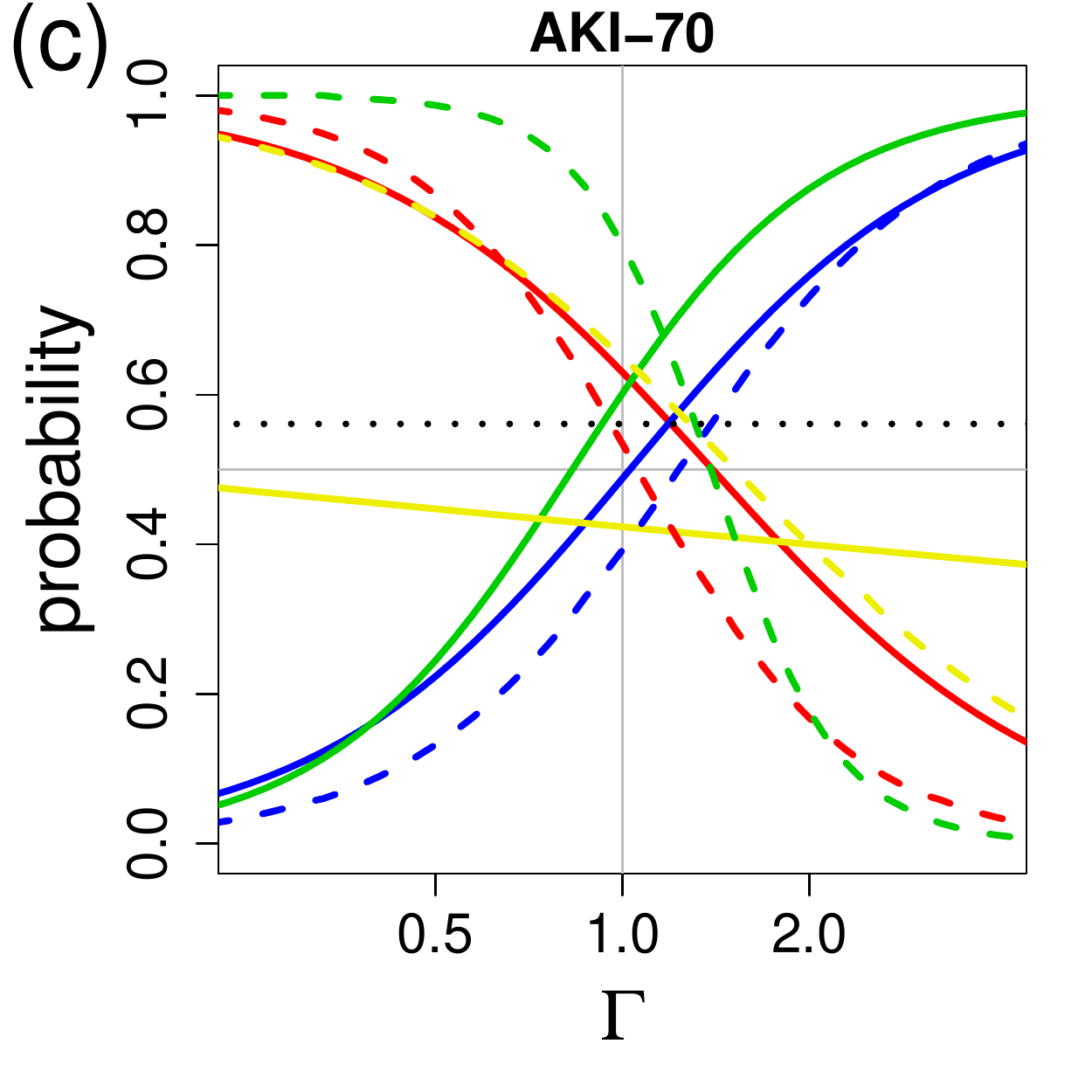}
\includegraphics[width=.242\textwidth]{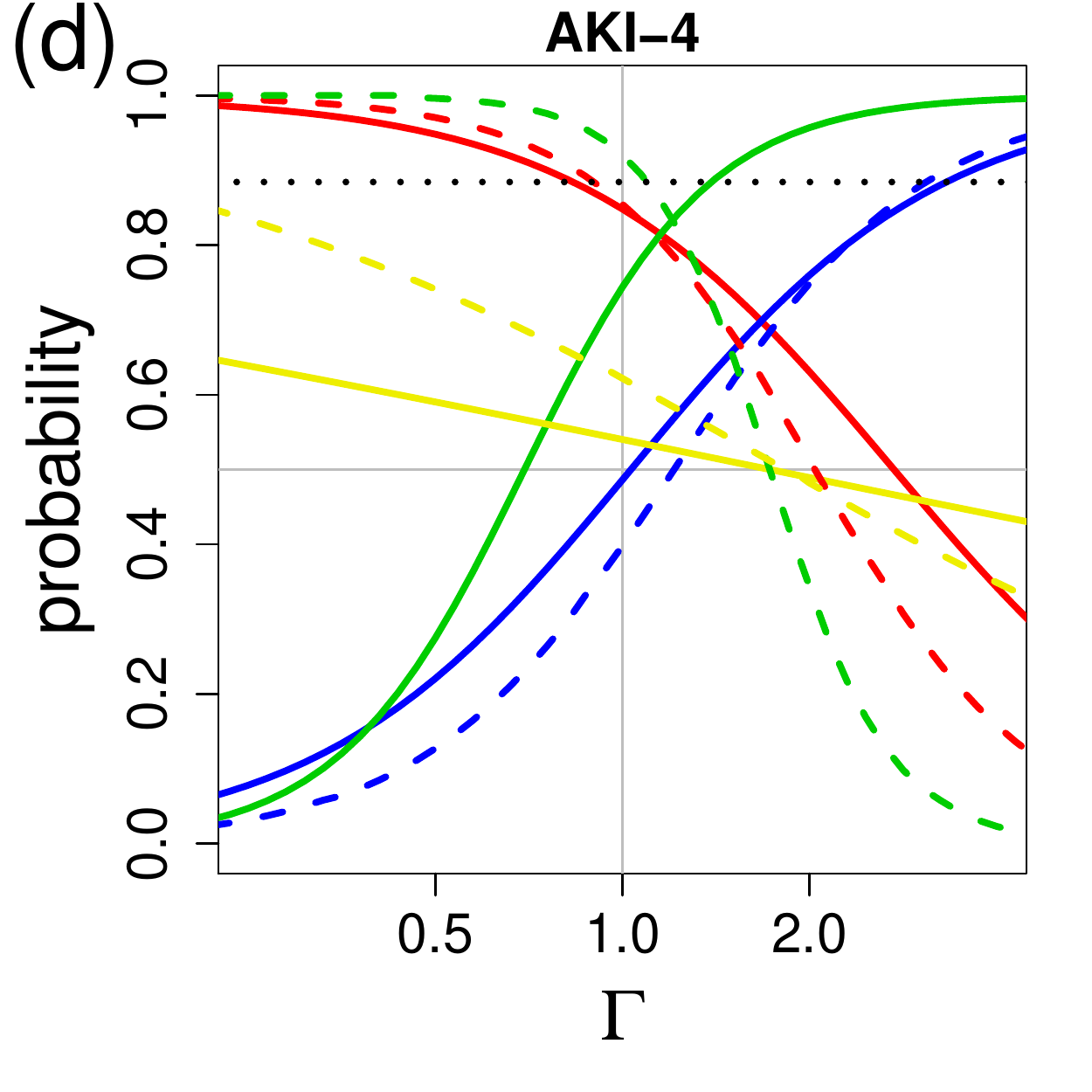}
\caption{\label{Prob_vs_gamma}The probability of AKI prediction as a function of the scale readjustment factor $\Gamma$ (see text).
The dashed lines correspond to f-SVM signatures and the solid lines to LASSO signatures. Colors 
indicate the underlying scaling methods of the signatures:  total spectral area (red), creatinine (blue), PQN (green), and TSP (yellow). The black dotted line gives the zero-sum predictions, which are independent of $\Gamma$.
}
\end{figure}

We discuss some misclassifications in more detail. 
Patients AKI-35 and AKI-106 developed severe AKI after 48 h but were not predicted to do so by two of 
the f-SVM signatures, namely those for creatinine and total spectral area scaled data.   
A key observation of this paper is that these misclassifications could be traced back to data normalization. In theory, the
predictions could have been saved by readjusting the scaling of these fingerprints (however, the necessary scaling factor is not evident \textit{a priori}).
Figure \ref{Prob_vs_gamma} shows the prediction probabilities of these signatures on the $y$-axis.
If this probability was above 0.5 we predicted an onset of AKI, otherwise we did not.
The $x$-axis shows possible multiplicative scale adjustments $\Gamma$. A value of $\Gamma=1$ indicates the actual 
scale used for prediction. Values of $\Gamma<1$ correspond to a down-scaling of all buckets by the factor $\Gamma$, 
while values of $\Gamma>1$ correspond to an up-scaling by $\Gamma$.
The plot thus shows the probability of developing AKI as a function of the scale-readjustment factor $\Gamma$.
The dashed lines correspond to f-SVM and the solid lines to LASSO signatures. Colors 
indicate the underlying scaling methods of the signatures:  
total spectral area (red), creatinine (blue), PQN (green), and TSP (yellow).

On the original scale ($\Gamma=1$) patient AKI-35 shows prediction probabilities below 0.5 for f-SVM-creatinine
(blue dashed line) and f-SVM-total-area (red dashed line). Let $k_{\textrm{crea}}$ be the ratio of the creatinine signal (from 3.055 to 3.013 ppm) and the remaining spectral area after exclusion of the D-mannitol region (4.0 to 3.5 ppm). For AKI-35 we observed $k_{\textrm{crea}}=0.069$, while the median over all AKI predicted samples was $k_{\textrm{crea}}=0.036$. In line with this, up-scaling the fingerprint rescues the prediction for this patient (blue dashed line in Figure \ref{Prob_vs_gamma}a). Similarly, $k_{\textrm{D-man}}$, the ratio between D-mannitol and the remaining spectral area, was $2.22$, while the median over all AKI predicted samples was 2.42, which explains why the f-SVM-total-area prediction could be rescued by a down-scaling of the fingerprint (red dashed line in Figure \ref{Prob_vs_gamma}a). Similar problems can be observed for patient AKI-106, and for the LASSO-creatinine signature in patients AKI-70 and AKI-4.

For the plasma data, we observed inconsistent predictions for 19\% (6\%) of the patients for f-SVM (LASSO) 
(Supplementary Figure S3). Here, no method identified the high-risk group completely, with best identification shown for LASSO on PQN data. 

In summary, we observed inconsistencies in the selection of biomarkers and more importantly in the prediction 
of patient outcome across the four scaling methods. These inconsistencies were most pronounced for the filter-based feature 
selection in f-SVM and affected urinary fingerprints more than plasma fingerprints. Still they could be observed for 
all analysis scenarios. For several patients a simple scale adjustment would have saved the prediction. 
The next section will show that zero-sum logistic regression resolves these inconsistencies fully and leads to more accurate predictions.

\subsection{Logistic zero-sum regression}
\label{LogZeroSum}
Zero-sum regression \cite{Linetal2014} is a novel machine-learning algorithm that is insensitive to rescaling the 
data \cite{Altenbuchingeretal2016}. It allows for a selection of biomarkers that does not depend on the units
chosen. The classifications of patients that result from these signatures do not
depend on any scaling of the data either. In fact, patients can be classified with spectral data that were not 
normalized at all.

For the reader's convenience we review the concept here in a nutshell:
Let $(x_i,y_i)$ be metabolomics data, where $x_{ij}$ is the logarithm of the intensity for bucket $j\in \{1,\ldots,p\}$ in
sample $i\in \{1,\ldots,N\}$ and $y_i$ the corresponding clinical response of patient $i$. In regression analysis the data sets need to be normalized to a 
common unit. Note that the data are on a logarithmic scale, therefore the scaling to a common unit becomes a shifting of the 
spectrum $x_{i}$ by some sample specific value $\gamma_i$.  
Thus for normalized data the regression equation reads
 \begin{equation}
y_i=\beta_0+\sum_{j=1}^p\beta_j (x_{ij}+\gamma_i) + \epsilon_{i}.
\label{regshift}
\end{equation}
Now note that the equation becomes independent of the normalization $\gamma_i$ if and only if the regression coefficients 
$\beta_j$ sum up to zero, i.e.,
\begin{equation}
\sum_{j=1}^p\beta_j=0\,.
\label{ZSum}
\end{equation}
This is the idea of zero-sum regression. In the machine-learning context of high-content data, zero-sum regression can be 
combined with LASSO or elastic-net regularization and shows predictive performances that were not compromised by the
zero-sum constraint \cite{Altenbuchingeretal2016}.
 
In many biomarker discovery challenges  the response $y$ is not continuous but binary. 
We thus need to extend the concept of zero-sum regression from linear regression to classification. We do this by introducing logistic zero-sum regression.  

In standard logistic regression the log-likelihood of normalized data reads
\begin{equation}
\label{LogLiLogist}
\mathcal L_0(\beta_0,\beta)=\frac{1}{N}\!\left\{\sum_{i=1}^Ny_i\! \left(\beta _0+\tilde x_i^T\beta \right)-\log\! \left(e^{\beta
   _0+\tilde x_i^T\beta }+1\right)\right\},
\end{equation}
where we abbreviated $\tilde x_{ij}=x_{ij}+\gamma_i$. In this generalized linear model the log-likelihood again becomes independent of the normalization $\gamma_i$ if
the regression coefficients $\beta_j$ add up to zero. This statement also holds if we add the penalizing term $\lambda P_{\alpha}(\beta)=\lambda(\alpha||\beta||_1+(1-\alpha)||\beta||_2)$, which corresponds to the elastic-net regularization penalty. The parameter $\lambda$ is calibrated in cross-validation \cite{Friedmanetal2010}, while $\alpha$ is usually fixed to a specific value in $[0,1]$. Here, $\alpha=0$ corresponds to ridge regression \cite{Hoerl1970} and $\alpha=1$ to the LASSO \cite{Tibshirani1996}.
Thus, logistic zero-sum regression amounts to finding coefficients $(\beta_0, \beta_j)$ that minimize
\begin{equation}
 -\mathcal L_0(\beta_0,\beta)+\lambda P_\alpha(\beta)\quad\mbox{subject to}\quad \sum_{j=1}^p\beta_j=0\,.
   \label{LogLiZSum}
\end{equation}
For all applications throughout the article, we have chosen $\alpha=1$.

As proposed by \cite{Friedmanetal2010} we use the quadratic approximation of Eq. (\ref{LogLiLogist}) locally
at the current parameters $(\tilde \beta_0, \tilde \beta)$ of the optimization algorithm:

\begin{align}
 l (\beta_0, \beta)& =- \frac{1}{2N}\sum_{i=1}^N
\Big( w_i
\left(z_i-\beta _0-x_i^T\beta\right)^2+C_i\Big)
\end{align}
with
\begin{align}
w_i& = \tilde p(x_i)(1-\tilde p(x_i))\,,  \\
z_i& = \tilde \beta _0 + x_i^T \tilde{\beta}
      +\frac{y_i-\tilde p(x_i)}{\tilde p(x_i)(1-\tilde p(x_i))}\,,\\
\tilde p(x_i) &= \frac{1}{1+e^{- \tilde\beta_0- x^T\tilde\beta}}\,.
\end{align}
$C_i$ is independent of $(\beta_0,\beta)$ and can thus be neglected. An efficient coordinate-descent algorithm
for zero-sum regression can be constructed by incorporating the zero-sum constraint 
into the log-likelihood by substituting $\beta_s$
with $\beta_s=-\sum_{\substack{j=1 \\ j\neq s}}^p \beta_j$ \cite{Altenbuchingeretal2016}. Calculating the partial derivative of the resulting log-likelihood 
with respect to $\beta_k$ ($k \neq s$) and solving for $\beta_k$ yields an update scheme for  $\beta_k$ and $\beta_s$:
 \begin{align}
      \hat\beta_k &= \frac{1}{a_{ks}} \cdot
      \begin{cases}
       b_{ks} - \lambda \alpha   &\text{if }  \hat\beta_k>0 \wedge \hat\beta_s > 0\,,\\
      ( b_{ks} - 2\lambda \alpha ) &\text{if }  \hat\beta_k>0 \wedge \hat\beta_s < 0\,,\\
      ( b_{ks} + 2\lambda \alpha ) &\text{if }  \hat\beta_k<0 \wedge \hat\beta_s > 0\,,\\
       b_{ks}    &\text{if }  \hat\beta_k<0 \wedge \hat\beta_s < 0\,,\\
      \text{else skip update}\,,
		      \end{cases}  \\
      a_{ks}=& \frac{1}{N}\sum_{i=1}^N  w_i( -x_{ik}+ x_{is} )^2  +2 \lambda 
		 (1-\alpha) \,,   \\
      b_{ks}=& \frac{1}{N}\sum_{i=1}^N w_i(x_{ik}- x_{is})\nonumber\\&\cdot
	  \Big( z_i - \beta_0 - \sum_{\substack{j=1 \\ j\neq s,k}}^P x_{ij}\beta_{j}
	  +  x_{is}\sum_{\substack{j=1 \\ j\neq s,k}}^p \beta_j  \Big)\nonumber\\&-
       \lambda
      (1-\alpha)  
       \sum_{\substack{j=1 \\ j\neq s,k}}^p\beta_j \,.
      \end{align}
The update value for $\beta_s$ is given by $\beta_s=-\sum_{\substack{j=1 \\ j\neq s}}^p \beta_j$.
After each update the approximation has to be renewed.
Since we update $\beta_k$ and $\beta_s$ simultaneously the number of updates scales quadratically with the 
number of features. Therefore, the minimization problem (\ref{LogLiZSum}) will be computationally demanding if $x$ becomes high-dimensional.

\subsection{HPC Implementation}
We performed a nested cross-validation (CV), where the inner CV uses 10 folds 
to determine a suitable regularization strength $\lambda$ and the outer CV is a leave-one-out CV 
to evaluate the prediction accuracy of the resulting models. 

We are evaluating two different data sets, each within a leave-one-out CV, one data set with 4 and the other with 3 different 
normalizations, and therefore we have to do 107$\times$4+86$\times$3=686 independent inner CVs. Each of these CVs is performed on an approximated regularization path of length 500, however, the algorithm stops when overfitting occurs. In total we performed $1,\!790,\!855$ model fits. This number presents an obvious computational challenge.
To avoid numerical uncertainties in the feature selection we used the very precise convergence criterion of $10^{-8}$, which additionally increases the computational effort.

On a standard server (2 Intel Xeon X5650 processors with 6 cores each), 
the complete calculation took about two days. We were able to bring the compute time down to 12 minutes by developing an HPC implementation of  \textit{zeroSum} and executing it on 174 nodes of the supercomputer QPACE 3 operated by the computational particle physics group. 

QPACE 3 currently comprises 352 Intel Xeon Phi 7210 (``Knights Landing'') processors, connected by an OmniPath network. Each processor contains 64 compute cores, and each of these compute cores contains two 512-bit wide vector units. To run efficiently on this machine, the \textit{zeroSum} C code was extended to use AVX512 vector intrinsics for the calculation of the coordinate-descent updates.
OpenMP is used to parallelize the inner CV so that the data has to be stored in memory only once and can be
accessed from all folds.

Additionally, we provide a new \textit{zeroSum} \textit{R} package, which is a wrapper around the HPC version
and can easily be used within \textit{R}. This \textit{R} package also includes functions for exporting and importing
all necessary files for/from the HPC implementation. 

For users without access to HPC facilities, our package can be run on a regular workstation with reduced convergence
precision and a shorter regularization path in less than one hour.

\begin{figure}[ht]
\includegraphics[width=.5\textwidth]{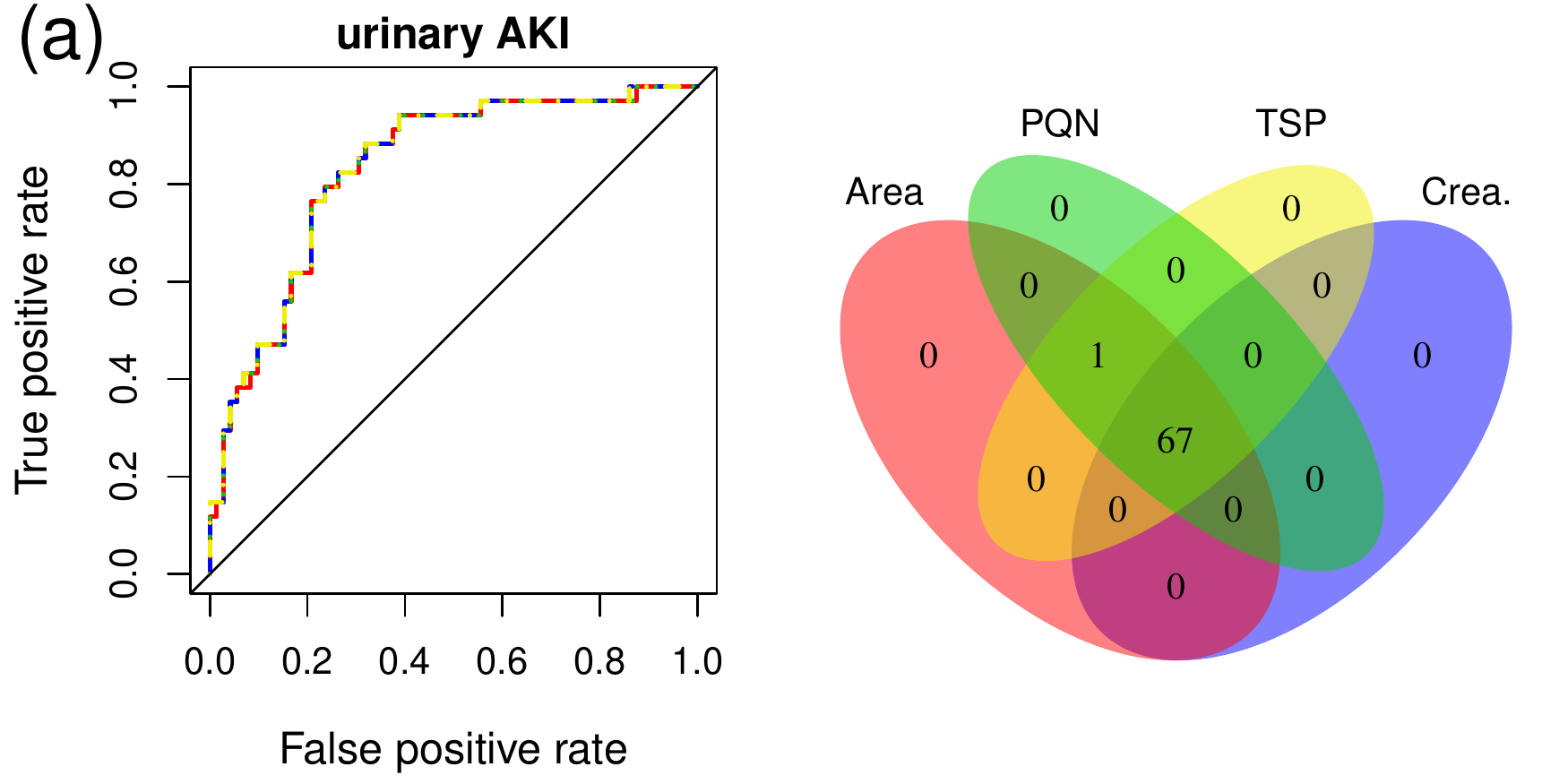}
\includegraphics[width=.5\textwidth]{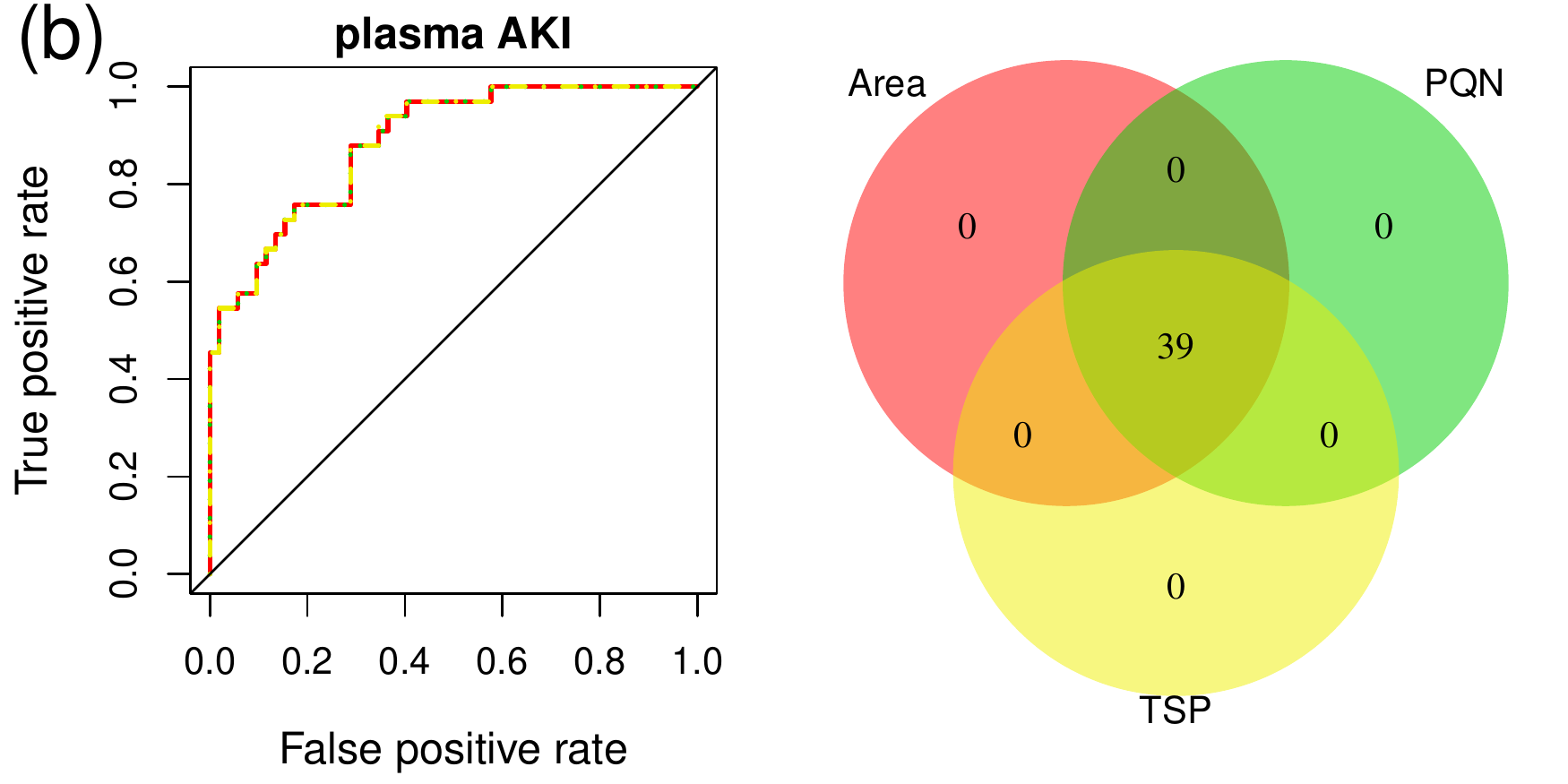}
\caption{\label{ROC_zeroSum} Receiver operating characteristic (ROC) curves for logistic zero-sum regression, after application of four different normalization strategies, 
scaling to total spectral area (red solid line), scaling to creatinine (blue dashed line), Probabilistic Quotient Normalization (PQN) (green dotted line), and scaling to TSP (yellow dashed-dotted line) for (a) the urinary AKI and (b) the plasma AKI data set. The right column shows the number of features included in the respective classification models in Venn diagrams. The corresponding models were built by averaging over all models of the outer CV loop. 
}
\end{figure}

\subsection{Logistic zero-sum regression is scale independent}

Logistic zero-sum regression is not affected by changes in scale. 
Actually, it will always yield the same result regardless of the scaling method used. Figure \ref{ROC_zeroSum} gives the corresponding ROC curves and Venn diagrams for the four normalization strategies investigated.
Since zero-sum models are scale independent, this classification approach always identified the same biomarkers and assigned the same weights to them independently of the normalization method chosen. Supplementary file 1 (supplementary file 2) lists all NMR buckets and corresponding metabolites for the urinary (plasma) data set with regression weights not equal to zero in at least one classification-normalization approach. Again consider the urinary AKI data set. Metabolites with large absolute regression weights ($\geq 0.2$) in the zero-sum models included both endo- as well as exogenous compounds. The largest positive regression weight was assigned to a bucket at 3.285 ppm, which was identified as a superposition of myo-inositol, taurine, 4-hydroxy-propofol-4-OH-D-glucuronide, and tentatively to D-glucuronic acid as well as an unknown metabolite. Other important buckets with large positive regression weights comprised propofol-glucuronide,  4-hydroxy-propofol-1-OH-D-glucuronide, isobutyrylglycine, broad protein signals, and probably 2-oxoisovaleric acid, indoxyl sulfate, as well as unidentified metabolites. Buckets with large negative regression weights correspond to a superposition of carnitine and 4-hydroxy-propofol-4-OH-D-glucuronide, N-acetyl-L-glutamine, and probably 4-hydroxyphenylacetic acid, as well as unidentified metabolites. Positive/negative regression coefficients indicate an up-/down-regulation of the corresponding metabolite in the AKI in comparison to the non-AKI group. Consequently, we can report an up-regulation of exogenous compounds such as glucuronide conjugates of propofol, an anesthetic agent which had been administered during the cardiac surgery and is mainly metabolized by glucuronidation \cite{Zachariasetal2015}. Elevated urinary levels of propofol conjugates in the AKI group point to a delayed excretion of exogenous compounds caused by reduced glomerular filtration or prolonged administration \cite{Zachariasetal2013a,Zachariasetal2015}. Higher absolute concentrations of propofol-glucuronide ($p$-value $= 0.02$) have already been reported for the 24 h urine NMR fingerprints in \cite{Zachariasetal2015}. An up-regulation of carnitine in the non-AKI group is indicated by its large negative regression weight. This observation has already been reported in \cite{Zachariasetal2013a}, where an up-regulation of carnitine, whose main function is the transport of long-chain fatty acids into the mitochondria for subsequent beta-oxidation \cite{Arduinietal2008}, in the non-AKI group has been discussed as a successful protective response against ischemic injury.

Also, logistic zero-sum regression returned the same predictions for all patients regardless of the applied normalization strategy, summarized in Figure \ref{classification_summary_AKI}, as well as in Supplementary Figure S3. Furthermore, the zero-sum constraint did not compromise predictive performance for the urinary and plasma AKI data set, respectively (Table \ref{ROC_table}, Figures \ref{ROC_zeroSum}a and \ref{ROC_zeroSum}b). AUC values ranged among the highest for  the urinary AKI data set, and yielded the largest value among all three classification methods in combination with three different normalization strategies for the plasma AKI data set. Most importantly, zero-sum regression reliably identified patients that developed severe AKI (stage 2 and 3, highlighted by yellow dashed area). Only LASSO on PQN data gave competitive results. With respect to AUC, zero-sum was still superior. Moreover, a sample-specific rescaling $\Gamma$ has no effect on the corresponding zero-sum prediction probabilities, as illustrated in Figure \ref{Prob_vs_gamma} by the black dotted lines.

\section{Discussion}

We first pointed out some intrinsic problems in metabolomics biomarker detection. The identification of biomarkers strongly depends on the method selected for scaling the reference profiles. For stand-alone biomarkers, accordance across scales can be as low as 0\%. The same is true for multivariate biomarker signatures when filtering approaches like filtering by the \textit{t}-statistics are used, while wrapping approaches like the LASSO yield somewhat more consistent yet still not scale-independent signatures. More importantly,
the prediction of a patient's outcome can change depending on the scaling method employed. In fact, for some cases we could attribute failing prediction to inappropriate
scaling of the raw data.  To overcome this problem we suggest logistic zero-sum regression, which provides a completely scale-independent analysis. It always selects the same biomarkers, assigns the same weights to them, and predicts the same outcome no matter whether the raw data was scaled relative to total spectral area, creatinine, TSP, any other
reference point, by PQN, or not at all. Furthermore, zero-sum regression was among the best AKI predictors both for urinary and plasma fingerprints. 
Also, the number of chosen features did not change significantly compared to other signatures, allowing for transfer of zero-sum signatures into clinical practice.

The interplay of statistical analysis and normalization protocols has been described by several authors 
\cite{Hochreinetal2015, Kohletal2012, Gromskietal2015,Jauhiainenetal2014, Saccenti2016}. 
Proposed strategies to overcome these issues included the selection of the most ``robust'' normalization  \cite{Hochreinetal2015} or the parallel application of several 
normalization strategies and a subsequent meta analysis of the individual results \cite{Saccenti2016}. 
Clearly, these strategies deal with the problem, but unlike zero-sum regression do not solve it.

The zero-sum signatures did not identify completely new biomarkers. In fact, most of the underlying metabolites of the signatures were described before.
However, we believe that zero-sum signatures make better use of these biomarkers by combining and weighing them differently. Predictions are no longer
confined to a specific normalization protocol but hold in general. Biological interpretation becomes easier as there are fewer user-dependent choices 
to make. It is a set of metabolites that hold predictive information and not their abundance relative to an arbitrarily  chosen reference point. 
Moreover, signatures can be validated across studies and tested in different labs even if scaling protocols do not match or if no scaling is applied at all. This fact has the potential to greatly enhance the reproducibility of clinical trials.

The urinary AKI data set is a good example for data where a proper normalization cannot replace independence of normalization. We believe that this data set just cannot 
be normalized properly: varying urine density precludes TSP normalization, the strong confounding of spectral areas by D-mannitol signals and by proteinuria 
eliminates both normalization to the total spectral area and PQN, while creatinine excretion is entangled with kidney function and varies strongly
across patients, disqualifying it from use as a common reference. Moreover, correcting for artifacts like D-mannitol requires that the existence of an artifact is known for a specific patient.

Interestingly, the LASSO-TSP signature worked surprisingly well in spite of strong variability in urinary density across patients. Coincidently, however, this signature had weights adding up to almost (yet not exactly) zero. 
LASSO is an algorithm used in artificial intelligence.  Here it detected the intrinsic normalization problems with the TSP-normalization and ``intelligently'' decided
for a scale-independent zero-sum type signature \cite{Altenbuchingeretal2016}.

Our \textit{zeroSum} package can be used on a desktop computer for small data sets and reduced precision. Large clinical studies
with several thousand patients require the use of an HPC infrastructure. We here offer code that can be used for such
studies.

In conclusion, we provide a high-performance classification framework independent of prior data normalization, which reduces the number of user-adjustable parameters and should be ideally suited for the transfer of metabolic signatures across labs. Furthermore, we expect that the results presented here also hold for metabolomics data generated by other methods such as mass spectrometry. 

\section{Acknowledgement}
The authors thank Drs. Gunnar Schley, Carsten Willam, and Kai-Uwe Eckardt for providing the urine and plasma specimens of the AKI data sets. Financial support:  e:Med initiative of the German Ministry for Education and Research (grant 031A428A) and the German Research Foundation (SFB/TRR-55).

\section{Conflicts of interests}
None.

\bibliographystyle{achemso}

\includepdf[pages=-]{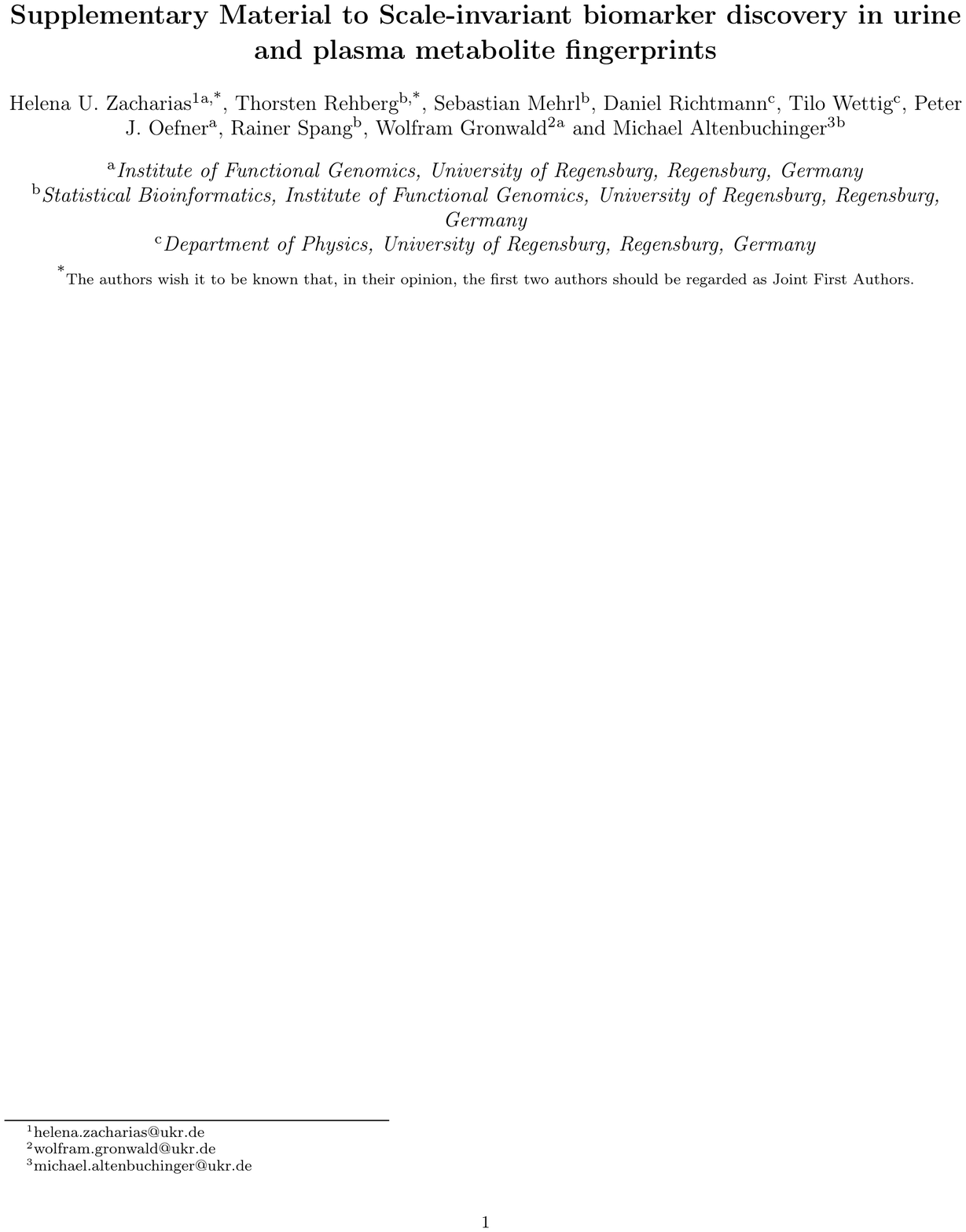}

\end{document}